\documentclass{article}
\usepackage[utf8]{inputenc}
\usepackage{authblk}
\providecommand{\keywords}[1]{\textbf{\textit{Keywords:}} #1}
\providecommand{\PACS}[1]{\textbf{PACS} #1}
\providecommand{\abbreviations}[1]{\textbf{Abbreviations} #1}
\providecommand{\conflictsofinterest}[1]{\textbf{Conflicts of interest:} #1}
\providecommand{\acknowledgments}[1]{\textbf{Acknowledgments:} #1}
\providecommand{\funding}[1]{\textbf{Funding:} #1}
\providecommand{\authorcontributions}[1]{\textbf{Author contributions:} #1}

\usepackage{amsmath}
\usepackage{graphicx}
\graphicspath{ {./images/} }
\usepackage{hyphenat}
\usepackage{latexsym}
\usepackage{stackrel,amssymb}
\usepackage{esint}

\title{Interferometer Euro-Asian Network:  Detection~Characteristics for Signals of Known Shape}
\author[1,2]{Valentin Rudenko \thanks{Corresponding Author:  valentin.rudenko@gmail.com}}
\author[3]{Svetlana Andrusenko}
\author[1,3]{ Daniil Krichevskiy}
\author[1,3]{Gevorg Manucharyan}
\affil[1]{Sternberg Astronomical Institute, Lomonosov Moscow State University}
\affil[2]{Faculty of Physics, Lomonosov Moscow State University}
\affil[3]{Department of Physics, Bauman Moscow State Technical University}
\date{}

\begin{document}

\maketitle

\begin{abstract}
    In this paper, we estimate efficiency of a conceivable Euro-Asian network of gravitational wave (GW) interferometers that might be realized having in mind a plan of construction of third generation interferometer in Novosibirsk region. Subsequently, some network would be composed, including four GW detectors. Among them there are the already active interferometers VIRGO (Italy) and KAGRA (Japan), Indian interferometer under construction---LIGO India and the interferometer in Siberia mentioned above. The quality of network in question is considered on the base of typical numerical criteria of efficiency for detecting GW signals of known structure---radiation of relativistic binary coalescence.
\end{abstract}



\section{Introduction}

In September 2015, the first recording of a gravitational-wave burst from the merger of a relativistic binary, whose components were evaluated as black holes (BH), took place. The detection of this event was carried out while using LIGO detectors \cite{Abbott061102}---large-scale laser interferometers with pendant-mirrors that play role of test masses of a gravitational gradiometer, which  these detectors essentially are. On the same arrangements until the end of 2017 other similar signals were detected \cite{Abbott241103,Abbott221101}. A qualitative step was the registration of GW170814 burst from the merger of BH binary ($M = 30\,{ M }_{ \odot } $) by three detectors (including a similar interferometer VIRGO in Europe) from the distance of $540 $ Mps \cite{Abbott141101}, which allowed reducing the localization zone of the source on the celestial sphere by an order of magnitude, up to $\sim$60 $deg^{2}$. Finally, a gravitational wave (GW) signal from neutron stars (NS) merger was registered, coinciding with GRB170817A gamma pulse (with $1.7$ s delay) \cite{Abbott161101}, confirming the hypothesis that short gamma bursts are born as a result of the NS-binary merger. All of these facts allow to claim confidently real occurrence of a new gravitational-wave channel of astrophysical information and heuristic value of multi-messenger astronomy, i.e., the strategy of parallel observation of transients on detectors of different physical nature.

In this situation, the task of creating an optimal terrestrial network of gravitational antennas both in terms of their geographical location and in terms of the nature of their interconnection---methods~of processing and decoding of their common signals against the background of local and global noises, is topical \cite{gus-ru}. The optimization of the geographical location of laser GW antennas on the globe (in the configuration of ``Einstein Telescope'' with axially symmetric (cylindrical) antenna power pattern~\cite{Sathyaprakash2012}) was carried out in the papers \cite{Raffai2013,Raffai2015} by numerical Monte Carlo simulation with Markov chains. Criteria~for optimization were the accuracy of estimates of the polarization of GW signals, the angular localization of the sources on the celestial sphere, and the parameters of so-called ``chirp'' signals. In~the light of these works at the current stage the prospect of a European-Asian network (EAN) of four antennas in the northern hemisphere: VIRGO in Italy, KAGRA in Japan, LIGO-India in India, and the planned new additional antenna in Novosibirsk was discussed \cite{Novosibirsk}. In order to assess scientific feasibility and efficiency of similar network the calculation of its main characteristics was performed in the approach developed in the articles  \cite{Raffai2013,Raffai2015}. In our paper, we consider real single interferometers (existing and under construction) as components of the network with the already given orientation, raising and solving the question about the optimal orientation of the planned detector in Novosibirsk (DN). For comparison, we will use two intercontinental global networks: LHVI (LIGO Livingston, LIGO Hanford, VIRGO, LIGO India), and LHVK (LIGO Livingston, LIGO Hanford, VIRGO, KAGRA). Table~\ref{tab:1} shows the coordinates of the gravity detectors that make up these networks. The detector orientation angle $ \gamma $ is defined as the angle between the southward direction at the detector location and the  bisector of the angle formed by its arms, measured counterclockwise.

\begin{table}
\caption{Detector data (from \cite{Arnaud2002, Akutsu2018, India2019}); all angles given in degrees.}
\label{tab:1}
\centering
\begin{tabular}{cccc}  
\textbf{Detector}	& \textbf{Latitude \boldmath{$\Phi $}}	& \textbf{Longitude \boldmath{$\lambda$}} & \textbf{Orientation \boldmath{$\gamma$}}\\
LIGO Hanford & 46.5 & 119.4 & 261.8\\
LIGO Livingston & 30.6 & 90.8 & 333.0\\
VIRGO & 43.6 & $-$10.5 & 206.5\\
KAGRA & 36.4 & $-$137.3 & 163.3\\
LIGO India & 19.6 & $-$77.0 & 254.0\\
Novosibirsk & 55.0  & $-$82.9 & to be defined\\
\end{tabular}
\end{table}

We perform a complete analysis of the proposed EAN network within the above criteria in order to select the optimal detector orientation angle of DN \cite{Raffai2013,Raffai2015}. On their basis, a generalized integral efficiency criterion is formed, the maximization of which by the orientation angle provides the most effective angle of DN.

It should be noted that, in this paper, we neglect anisotropic distribution of sources in the celestial sphere. This approximation works well if the horizon distances of detectors in question are of the order of 200--300 $Mpc$, because on this scale the Universe can be considered as homogeneous and isotropic with high accuracy. The sensitivity of modern LIGO, VIRGO detectors makes it possible to work in this approximation.

\section{Network Power Pattern}
\label{sct:2}

The concept of "a network of ground-based detectors" is inextricably linked to the task of constructing a network power pattern of its components. Let us review the basic information necessary for the construction of the pattern.

\begin{figure}
\centering
\includegraphics[width=7 cm]{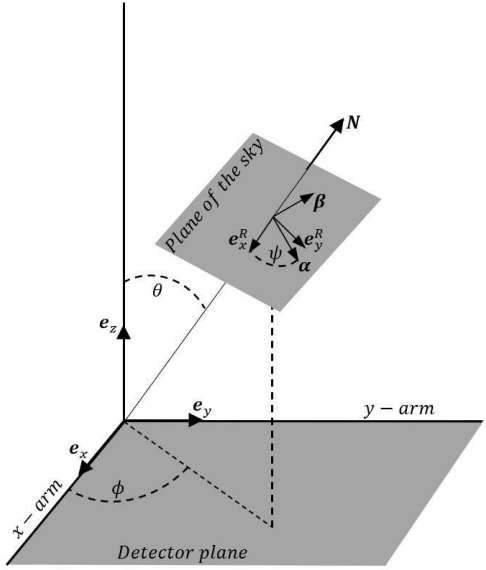}
\caption{The relative orientation of the celestial and detector frames. Reproduction from \cite{Schutz2009}.} 
\label{fig:1}
\end{figure}

In the frame of the GW propagating along the $z$ axis, the perturbation of the metric in the TT-gauge looks like:
\begin{equation}
    { h }_{ \mu \nu  }=\begin{pmatrix} 0 & 0 & 0 & 0 \\ 0 & { h }_{ + }(t) & { h }_{ \times  }(t) & 0 \\ 0 & { h }_{ \times  }(t) & -{ h }_{ + }(t) & 0 \\ 0 & 0 & 0 & 0 \end{pmatrix}.
\end{equation}
In the long wavelength approximation (the GW wavelength is much larger than the interferometer arm length), the detector response can be evaluated as
\begin{equation}
h(t)=\frac { \delta L }{ L } ={ F }_{ + }(\theta,\phi,\psi ){ h }_{ + }(t)+{ F }_{ \times }(\theta,\phi,\psi ){ h }_{ \times }(t),
\end{equation}
where ${ F }_{ + }(\theta, \phi, \psi )$, ${ F }_{ \times }(\theta, \phi, \psi )$ are the antenna pattern functions for the two polarizations, which are functions of the polar angle  $\theta$ and the azimuth angle $\phi$ of the spherical coordinate system (XY is the detector plane) and the polarization angle of the GW $\psi$. Figure~\ref{fig:1} shows the relative orientation of the celestial and detector frames. 

\begin{figure}
\centering
\includegraphics[width=5 cm]{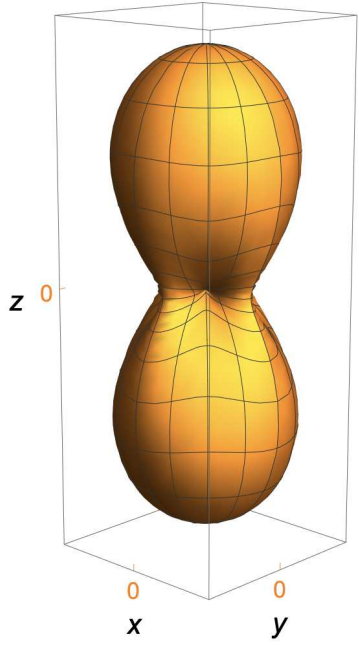}
\caption{Antenna power pattern of a single detector.} 
\label{fig:2}
\end{figure}

Formula (2) is valid in the coordinate frame which basis vectors coincide with the direction of the detector arms. Antenna pattern functions have the following form:
\begin{equation}
\begin{gathered}
{ F }_{ + }=\frac { 1 }{ 2 } (1+\cos ^{ 2 }{ \theta } )\cos { 2\phi } \cos { 2\psi } -\cos { \theta } \sin { 2\phi } \sin { 2\psi } , \\
{ F }_{ \times }=\frac { 1 }{ 2 } (1+\cos ^{ 2 }{ \theta } )\cos { 2\phi } \sin { 2\psi } +\cos { \theta } \sin { 2\phi } \cos { 2\psi }.
\end{gathered}
\end{equation}
If we consider a source that emits Fourier wave components $\tilde { { h } }_{+}(f)$, $\tilde { { h } }_{\times}(f)$, then the mean power SNR (signal-to-noise ratio) \cite{Schutz2011} is given by
\begin{equation}
    \left< { \rho  }^{ 2 } \right> =2P(\theta ,\phi )\int _{ 0 }^{ \infty  }{ \frac { { \left| \tilde { { h } }(f) \right|  }^{ 2 } }{ { S }_{ n }(f) } df }, 
\end{equation}
where  $P(\theta,\phi)$---antenna power pattern (Figure~\ref{fig:2}) which does not depend on $\psi$ :
\begin{equation}
  P(\theta, \phi )={ { F }_{ + }^{ 2 }(\theta, \phi, \psi ) }+{ { F }_{ \times }^{ 2 }(\theta, \phi, \psi ) },
\end{equation}
and $ { \left|\tilde { { h }}(f) \right| }^{ 2 }={ \left|\tilde { { h } }_{+}(f) \right| }^{ 2 }+{ \left| \tilde { { h } }_{\times}(f) \right| }^{ 2 } $. 

If the equatorial coordinate system is used Equation (3) can be rewritten using the matrix formalism. A detailed discussion of explicit formulas for $ { F }_{ + }$ and $ { F }_{ \times }$ can be found, for example, in \cite{JaranowskiSchutz1998,Bonazolla1996,Arnaud2002}. These formulas are not the same in cited papers, because of different definitions of used angles. We used those ones from \cite{Arnaud2002}, where source position is defined in the equatorial coordinate system by two angles: right ascension $ \alpha \in [-\pi ;\pi ] $ and declination $ \delta \in [-\frac { \pi }{ 2 } ;\frac { \pi }{ 2 }] $. Finally, one has
\begin{equation}
    \begin{pmatrix} { F }_{ + }(t) \\ { F }_{ \times }(t) \end{pmatrix}=\sin { \chi } \begin{pmatrix} \cos { 2\psi } & \sin { 2\psi }  \\ -\sin { 2\psi } & \cos { 2\psi }  \end{pmatrix}\begin{pmatrix} a(t) \\ b(t) \end{pmatrix},
\end{equation}

\begin{equation}
\begin{gathered}
  a(t)=-\frac { 1 }{ 16 } \sin { 2\gamma  } (3-\cos { 2\Phi } )(3-\cos { 2\delta  } )\cos { 2H(t) } -\\ \frac { 1 }{ 4 } \cos { 2\gamma  } \sin { \Phi } (3-\cos { 2\delta  } )\sin { 2H(t) } -\\ \frac { 1 }{ 4 } \sin { 2\gamma  } \sin { 2\Phi } \sin { 2\delta  } \cos { H(t) } -\frac { 1 }{ 2 } \cos { 2\gamma  } \cos { \Phi } \sin { H(t) } - \\\frac { 3 }{ 4 } \cos { 2\gamma  } \cos ^{ 2 }{ \Phi } \cos ^{ 2 }{ \delta  }, \\
  b(t)=-\cos { 2\gamma  } \sin { \Phi } \sin { \delta  } \sin { 2H(t) } +\\ \frac { 1 }{ 4 } \sin { 2\gamma  } (3-\cos { 2\Phi } )\sin { \delta  } \sin { 2H(t) } -\\\cos { 2\gamma  } \cos { \Phi } \cos { \delta  } \cos { H(t) } +\frac { 1 }{ 2 } \sin { 2\gamma  } \sin { 2\Phi } \cos { \delta  } \sin{H(t)},
\end{gathered}
\end{equation}

\begin{equation}
   H(t)=\omega t-(\alpha +\lambda)+{ T }_{ G }(0),
\end{equation}
where $\omega$ is the Earth's angular frequency around its axis, ${ T }_{ G }$ is sidereal time at $0$ h UT, $\chi$ is the opening angle of the interferometer arms, which is 90 degrees in our paper, but can also be 60 degrees for third-generation GW detector ``Einstein Telescope'' \cite{Freise2009}. The moment of time is chosen, so that $\omega t_{ 0 }+{ T}_{ G }(0)=0$.

In \cite{Finn2001}, it is shown that for a system of $N$ detectors network antenna power pattern ${ P }^{ N }$ is given~by:

\begin{equation}
    { P }^{ N }=\sum _{ k=1 }^{ 4 }{({ F }_{ +,k }^{ 2 }+{ F }_{ \times,k }^{ 2 }) }.
\end{equation}

\section{Parameters of Signals from Mergers}
\label{sct:3}

By now, only astrophysical GW bursts accompanying the merger of the relativistic binary components at the end of its evolution have been successfully detected. The form of the gravitational signal has been analyzed in detail in the literature and it is included in monographs. In particular, one of the first is the monograph \cite{MKW}, where the calculation of the spiral phase of the binary was made in the Newtonian approximation, and the rotational energy losses were calculated using the quadrupole Einstein formula for gravitational radiation. 

For the purposes of this paper, a rather simplified Newtonian form of gravitational chirp signal, which does not take into account post-Newtonian corrections is sufficient. It is a type of quasi-harmonic oscillation with increasing amplitude and carrier frequency \cite{Finn1993}. For small redshift $z\sim 0.1$, it can be written as
\begin{equation} 
h(t)=\frac { 4 }{ { d }_{ L } }\sqrt { { F }_{ + }^{ 2 }+{ F }_{ \times  }^{ 2 } } \frac { { G }^{ { 5 }/{ 3 } } }{ { c }^{ 4 } } {\mathcal{M} }^{ { 5 }/{ 3 } }{ \left( \pi f \right)  }^{ { 2 }/{ 3 } }cos\left(  { \Phi  }_{ c } +\Psi  \right), 
\end{equation}

\begin{equation}
\mathcal{M}={ \mu  }^{ { 3 }/{ 5 } }{ M }^{ { 2 }/{ 5 } },
\end{equation}

\begin{equation}
{ \Phi  }_{c} =-2{ \left( \frac { { c }^{ 3 }(T-t) }{ 5G{ M } }  \right)  }^{ { 5 }/{ 8 } },
\end{equation}

\begin{equation}
f=\frac { 1 }{ \pi { M } } { \left[ \frac { 5 }{ 256 } \frac { { { c }^{ 5 }M } }{ G(T-t) }  \right]  }^{ { 3 }/{ 8 } },
\end{equation}
where $\mathcal{M}$ is the chirp mass of a system, ${ d }_{ L }$ is the distance to the source, ${ \Phi  }_{c}$ is the phase, $f$ is the frequency, $\Psi$ is the initial phase, $M$ and $\mu$ is the total and reduced masses of the binary, and $T$ is the moment of coalescence.

From the very beginning of the search for GW bursts (or pulsed GW signals) of extraterrestrial origin, it was clear that the most likely prediction is to register the radiation that is generated by the cosmic catastrophe merger of relativistic (super-dense) binary stars \cite{BZR}. The structure of this GW impulse was evident from the physical picture of the process, the probability of occurrence was estimated by the data of population and lifetime of the binary systems in the galaxy, a fairly accurate calculation of the parameters of the burst was obtained from Newton's theory. In other words, it was the most reliable source of GW radiation to stimulate and develop very expensive GW observatories. Other~types of sources contained much more elements of uncertainty and, consequently, fewer guarantees for their successful registration. These a priori expectations were brilliantly confirmed in practice after the first registration of a chirp signal from a BH merging binary in 2015. Today, the roughly simple, but in detail, complex structure of  bursts like this generates intriguing possibilities of rapid progress of GW astronomy---through inverse problem of reconstruction of physics of relativistic objects by structure of emitted GW radiation. In this respect, after the success of the first registration, the problems of precision evaluation of fine details and parameters of the chirp come to the fore: the chirp mass of the system $\mathcal{M}$ and the distance to the source ${ d }_{ L }$, taking the relativistic corrections into~account.

As it will be shown below, the ability to evaluate the parameters of the chirp is one of the important criteria of the quality of created networks of GW detectors.  It is desirable to have not only the temporal structure of the signal (10), but also its amplitude and frequency-phase spectrum.

To conclude, we consider a Fourier image of signal (10), which is obtained in the approximation of the stationary phase \cite{Finn1993} and it is valid for frequencies up to $\sim$1500 Hz \cite{Cutler1994}:
\begin{equation}
\begin{gathered}
\tilde { h } (f)=\frac { 4 }{ { d }_{ L } } \frac { { G }^{ { 5 }/{ 6 } } }{ { c }^{ { 3 }/{ 2 } } }\sqrt { { F }_{ + }^{ 2 }+{ F }_{ \times  }^{ 2 } } { \left( \pi f \right)  }^{ -{ 7 }/{ 6 } }\sqrt { \frac { 5\pi  }{ 384 }  } { \mathcal{M} }^{ { 5 }/{ 6 } }  \\ exp\left( i\left[ 2\pi fT+\frac { 3 }{ 128 } { \left( \pi f\mathcal{M} \right)  }^{ { -5 }/{ 3 } }-\Psi +\frac { \pi  }{ 4 }  \right]  \right). 
\end{gathered}
\end{equation}

\section{Criteria of a Network}
\label{sct:4}

We use three independent criteria presented in the works~\cite{Raffai2013,Raffai2015} in order to choose the optimal angle of DN. These three conditions form an integral criterion, which is to be maximized by changing the orientation of DN, in order to find the most effective angle.

\subsection{Polarization Criterion I}
\unskip
An important feature a GW detectors network is its ability to assess the polarization of the received GW. This parameter is very important in solving the inverse problem of GW astronomy---the~construction of an astrophysical model of the source. The logic of introducing the corresponding criterion, the so-called “polarization criterion I”, is as follows. 
 
The strain output $ h $ of a detector (response to a passing GW with two polarizations ${ h }_{ + } $ and ${ h }_{ \times}$) is given by the formula (2), where $ { F }_{ + }$ and $ { F }_{ \times }$ are antenna patterns that correspond to two characteristic orientations of the interferometer. ${ F }_{ + }$ and $ { F }_{ \times }$ are the functions of the source coordinates on the celestial sphere and the polarization angle of the gravitational wave $\psi $.
 
Following \cite{Raffai2013}, we define $+$ and $\times$ integral functions for a network of four detectors:
\begin{equation}
{ F }^{ N }=\frac { 1 }{ 2 } \sqrt { { F }_{ 1 }^{ 2 }+{ F }_{ 2 }^{ 2 }+{ F }_{ 3 }^{ 2 }+{ F }_{ 4 }^{ 2 } } ,
\end{equation}
where $N$---stands for a network function, i.e., standing under the root ${F}_{1}...{F}_{4}$ all either correspond to the $+$ or $\times$ polarization. Obviously ${F }^{ N }$ depend on the polarization angle of $\psi $. 

To calculate criterion I, one needs to choose a system of a certain (preferred) polarization angle---dominant polarization frame (DPF) \cite{Klimenko2011}. In DPF, for each point on the celestial sphere $(\alpha ; \delta)$, we select a polarization angle (the angle by which source's natural polarization basis is rotated relative to the equatorial coordinate system basis) that maximizes the network factor ${ F }_{+}^{ N }$ and minimizes ${ F }_{\times}^{ N }$.  Consequently, for this direction $(\alpha ; \delta)$, the condition ${ F }_{+}^{ N } \ge { F }_{\times}^{ N } $ is valid. A detailed mathematical proof of that fact can be found in \cite{Drago2010}. For simplicity, we consider all of the detectors to have the same noise characteristics, which makes DPF easy to use. In general, when constructing DPF, one should consider the difference in noise background between interferometers.

While giving preference to one of the polarizations in DPF, the condition of approximate equality of factors ${ F }_{\times}^{N}$ and $ { F }_{+}^{N}$ has to be kept, i.e., $\frac {{ F }_{\times}^{ N }  }{ { F }_{+}^{ N } }\approx 1 $. This means that the gravity detector network will be sensitive to both gravity wave polarizations. It follows that a minimum difference of $\left| { F }_{\times}^{ N } - { F }_{+}^{ N }  \right| $ should be sought for all $(\alpha ; \delta)$. This leads to the quantitative formulation of the polarization criterion~I~\cite{Raffai2013}, which is to find the maximum of the functional: 
\begin{equation}
I={ (\frac { 1 }{ 4\pi } \oiint { { \left| { F }_{ + }^{ N }(\alpha ; \delta)-{ F }_{ \times }^{ N }(\alpha ; \delta) \right| }^{ 2 }d\Omega } ) }^{ -1/2 },
\end{equation}
where averaging of $\left| { F }_{\times}^{ N } - { F }_{+}^{ N }  \right| $ over celestial sphere takes place ($d\Omega$---solid angle).

\subsection{Localization Criterion D}

The following criterion of quality of a network of ground detectors is connected with its ability to define the angular position of a source. In astrometry, the classical problem of localization on celestial sphere of a radiation source (not too remote) is solved by a method of triangulation. 

The accuracy of triangulation naturally depends on the signal-to-noise ratio, which, in turn, is determined by the time spent on the measurement. Thus, it is important how long the source is in the sight of all detectors in the network. For GW transients, this time cannot be long (one deals with a single pulse source). Criterion $D$, however, characterizes the relative ability of networks of different configurations to determine the angular position of the source. An absolute evaluation of the degree of localization is not required here and the signal-to-noise ratio can be disregarded.

The strict formulation of criterion $D$ in \cite{Raffai2015} is based on statistical analysis \cite{Fairhurst2011} and it is associated with the functional 
\begin{equation}
    D={ \frac { 1 }{ 4\pi  } \oiint _{  }^{  }{ H(S-{ A }_{ 90 }(\alpha ,\delta ))d\Omega  }  },
\end{equation}
where $ H(x)$ is a Heaviside function and $ A_{0.90}$ is $90\%$ confidence localization region for a source located at sky position $(\alpha,\delta )$, $S$ is a preset threshold. Thus, to use this formula, we need to introduce a hypothesis about a priori localization zone statistics, which makes the task much more complicated. In~this regard, a simplified method of criterion D estimation proposed in \cite{Raffai2013} is considered below, which,~as verified in \cite{Raffai2015}, gives results that are close to the strict approach. 

Triangulation is based on the difference in time between the registration of signals by network detectors. The further apart the detectors are, the greater the time delay. To maximize the source location accuracy on the celestial sphere, the telescopes should be placed as far apart from each other as possible. According to \cite{Raffai2013}, for a network of four detectors, the $D$ criterion parameter is the area of the triangle formed by the three detectors in the network, which has the largest area of all possible detector combinations. Linear algebra formulas are sufficient in the calculations. If $O$---the center of the Earth, $A$, $B$, $C$---points where the detectors are located, the area of the corresponding triangle:
\begin{equation}
S_{ABC}=1/2|[\overrightarrow{AC},\overrightarrow{AB}]|=1/2|[\overrightarrow{OC}-\overrightarrow{OA};\overrightarrow{OB}-\overrightarrow{OA}]|.
\end{equation}

\subsection{ Parameters Reconstruction Criterion R }

The third criterion of network quality characterizes the possibility of reconstruction the parameters of the signal of known structures. For the EAN network, this criterion is calculated for the signals that accompany the merge of relativistic binary stars, i.e., chirp signals (see Section~\ref{sct:3}). It is worth noting that, in terms of gravitational radiation, merging binaries can be considered as ``standard sirens”, due~to the high accuracy of the intensity and shape of the emitted signal, as well as the practical absence of the effects of absorption and scattering of GW. The astrophysical characteristics of the merging binary can be determined by the intensity and shape of the recorded chirp signal. The main interesting parameters of this signal are the distance to the source  ${ d }_{ L }$ and the mass of the chirp $\mathcal{M}$. Interferometer~detects radiation from binary systems to within its noise background. The spectral noise density of LIGO interferometers has been repeatedly presented in documents and publications. In particular, when~calculating the $R$-criterion, we used sensitivity curves that were taken from the open LIGO data for O1 run \cite{O1data}.

According to the Maximum Likelihood Estimation in the additive Gaussian noise background model, the parameters of the received signal are evaluated by the Rao--Cramer bound. According to this criterion, the best possible estimates are obtained while using the Fisher information matrix~$ { \Gamma }_{ \alpha \beta }$~\cite{Kocsis2007}, in accordance with the formula
\begin{equation}
{ \Gamma }_{ \alpha \beta }=Re\left\{ 4\int _{ { f }_{ min } }^{ { f }_{ max } }{ \frac { \overline { { \partial }_{ \alpha }\tilde { h } (f) } { \partial }_{ \beta }\tilde { h } (f) }{ { S }_{ n }(f) } df } \right\},
\end{equation}
where $\tilde { h } (t)$  is the Fourier image of response of the detector (for the chirp signal, formula (14)), the~line above the Fourier image of response represents the complex conjugate, and ${S }_{ n}(f)$ is the spectral noise~density.

Rao--Cramer bound determines the best possible accuracy of chirp mass estimation \cite{Kocsis2007}:
\begin{equation}
{ \delta \mathcal{M} }^{ 2 }={ \left( { \Gamma }_{N}^{ -1 } \right) }_{\mathcal{M}\mathcal{M}},
\end{equation}
where ${ \Gamma }_{N}=\sum _{ i=1 }^{ N }{ { \Gamma }_{ i }^{ } } $, i.e., the Fisher information matrix for detector network, is the sum of the corresponding detector matrices constituting the network. It was shown in \cite{Raffai2013} that non-diagonal elements of matrix ${ \Gamma }_{N}$ have practically no influence on the error calculation, so they can be neglected in the calculation (19). The inverse value of the celestial-averaged relative error ${ < \frac { \delta \mathcal{M} }{ \mathcal{M} } > }^{ -1 }$, in fact, is a numerical expression of criterion $R$, which evaluates the ability of the network to reconstruct an important parameter of the astrophysical object---the chirp mass:
\begin{equation}
R={ \left( \frac { 1 }{ 4\pi } \oiint { { \left( \frac { \delta \mathcal{M} }{ \mathcal{M} } \right) }^{ 2 }d\Omega } \right) }^{ { -1 }/{ 2 } }={ \left< \frac { \delta \mathcal{M} }{ \mathcal{M} } \right> }^{ -1 },
\end{equation}
Consequently, the maximization of criterion $ R $ leads to the minimum relative error averaged over the celestial sphere in the estimation of chirp mass.

It is worth noting that signal (10) is defined by the chirp mass $\mathcal{M}$ and the distance to the source $d_{L}$, but only one criterion corresponding to reconstructing of parameters is built. It is due to the fact that a criterion based on $d_{L}$ gives almost the same results when one compares different networks~\cite{Raffai2013}. 

\subsection{ Integral Criterion C }

The last criterion proposed in \cite{Raffai2013} allows us to evaluate and compare different detector network~configurations.

The above criteria $ I $, $ D $, and $ R $ together define a three-dimensional space that can be used to define a point describing this configuration for any set of four detectors. Our goal is to find a configuration of EAN network detectors that matches these $ I $, $ D$ , and $ R $, so that they are maximal. These maximum values will be denoted as $I_{max}, D_{max}$ and $R_{max}$.

In this paper, we consider that the weight of each criterion is the same and therefore the values are normalized to the maximum values $I_{max}, D_{max}$ and $R_{max}$, which makes the value of criterion $C$ dimensionless. Thus:
\begin{equation}
C=\sqrt{(\frac{I}{I_{max}})^2+(\frac{D}{D_{max}})^2+(\frac{R}{R_{max}})^2}
\end{equation}

\section{Numerical Results and Conclusions}
\label{sct:5}

The main purpose of our study was to analyze the Euro-Asian GW interferometer network (EAN), taking into account the planned creation of such an interferometer in Siberia near Novosibirsk. At~the same time, a practical issue requiring calculations was the optimal orientation of such a detector, which would ensure maximum efficiency of the network as a whole. For the numerical evaluation of efficiency, we have used the quality criteria of GW networks, proposed in a number of previous works~\cite{Raffai2013, Raffai2015}. The logic of these criteria was briefly reproduced by us in Section~\ref{sct:4} and the calculation formulas (16), (18), (21) and (22) were given. The use of these criteria required numerical simulations of the network (EAN) and, above all, the reproduction of its antenna power pattern.

One of its main characteristics, such as proportion of the sky coverage areas (celestial sphere zones visible to all network detectors), is mainly determined by the ground-based configuration of detectors, which is given in our task. The sky coverage areas are zones for which the condition
\begin{equation}
    { P }^{ N }(\alpha ,\delta )\ge \frac { 1 }{ 2 } { P }_{ max }^{ N },
\end{equation}
is valid, where the ${ P }_{ max }^{ N }$ is the maximum value of the network antenna power pattern for that network. The results of celestial coverage calculation for European-Asian network (EAN), HLVK, and HLVI are shown in Table~\ref{tab:2}. 
\begin{table}
\caption{Effective reception area of the GW signal.}
\label{tab:2}
\centering
\begin{tabular}{cc}  
\textbf{Network} & \textbf{Effective Area}\\
EAN & 56\% \\
LHVK & 97\% \\
LHVI & 78\%  \\
\end{tabular}
\end{table}
One can see where the EAN is inferior to the other two networks. The larger the effective reception zone is, the further away the network detectors are from each other. The smaller the distance between detectors is, the better the network receives the signal from specific areas of the celestial sphere, but~worse on average. It follows that the use of an EAN network is prospective for receiving signals from certain areas of the celestial sphere.

The full set of quality criteria includes the evaluation of the parameters of the received signal. In~this way, the complete set can only be formulated for a specific type of GW source. As such, in this paper, we take the GW burst that accompanies a cosmic disaster---the merger of relativistic binaries. So~far,~this is the only astrophysical type of sources from which it was possible to register GW---signals, the so-called chirps from cosmological distances of hundreds of $Mpc$.

Numerical modeling was associated with code writing, which we performed in MATLAB language (version R2020a Update 3). Formula (7) were used to generate arrays of antenna power functions $(\alpha,\delta)$ for each detector from Table 1. As a check of the formulas correctness, the antenna power pattern we calculated for VIRGO was compared with that presented in \cite{Arnaud2002}. Adequacy of understanding of the DPF concept was confirmed by correct reproduction of $F_{\times}^{N}/F_{+}^{N}$ ratio for AHLV network (planned Australian detector near Perth, LIGO Hanford, LIGO Livingston, VIRGO) given in~\cite{LIGOReport2010}.

Numerical integration of surface integrals (16), (21) on a uniform grid $(\alpha,\delta)$ was carried out. The~integration of the Fisher information matrix (19) was performed on an irregular grid in frequencies, induced by the given curve of spectral noise density that is presented as an array of 3000 points \cite{O1data}.

As a characteristic source, as in \cite{Raffai2013}, we chose a binary neutron star with masses $1.4\,{ M }_{ \small{\bigodot} }$, without~spins, located at a distance of $1$ $Gpc$ from the Earth and with an orbital plane perpendicular to the line-of-sight.

Therefore, the main task was to find the optimal angle of orientation of DN as part of the EAN;~this~angle was the degree of freedom in the integration of parameters. Since the detector orientation angles, which differ from each other at ${ 90 }^{ \circ  } $, are equivalent, we calculated criteria for the EAN in the range ${ \gamma  }_{ Nsk }\in (0,\frac { \pi  }{ 2 } ]$ with step of ${ 1 }^{ \circ} $.

\begin{figure}
\centering
\includegraphics[width=9 cm]{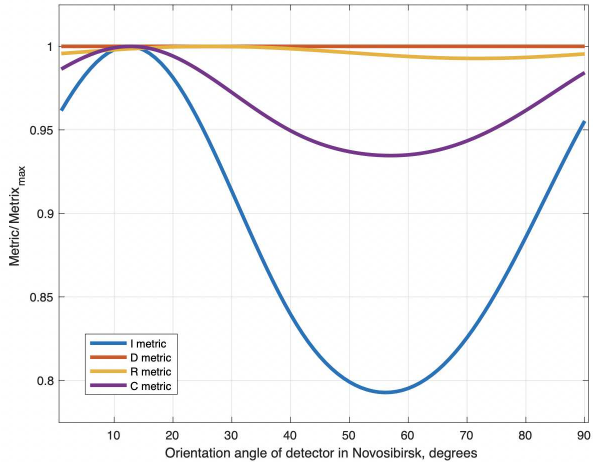}
\caption{Dependence of all criteria on orientation angle of detector in Novosibirsk.}
\label{fig:3}
\end{figure}

Figure~\ref{fig:3} shows the modeling results. Criterion $ D $ in our approximation does not depend on the orientation angle of DN. Criterion $ I $ is the most sensitive to changes of orientation angle and reaches its maximum at the angle ${ 12 }^{ \circ  }$. Criterion $R$ also depends on the orientation angle of DN, but is not so sensitive to its variations as $I$. $R$ reaches it maximum at ${ 25 }^{ \circ  }$. Integral criterion $ C $ reaches its maximum value at almost the same point as $I$ maximum and the optimal angle is ${ \gamma  }_{ Nsk }^{max}={ 13 }^{ \circ  } $.

\begin{figure}
\centering
\includegraphics[width=7 cm]{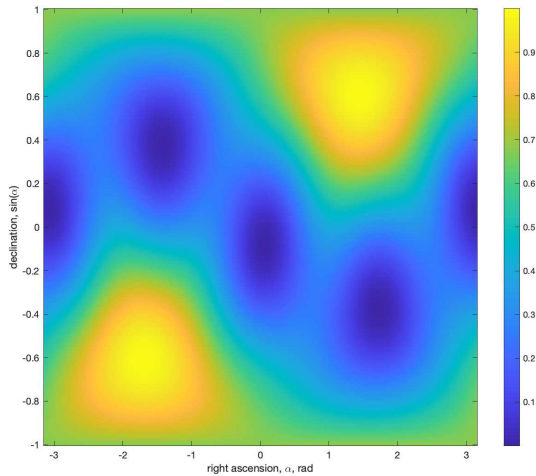}
\caption{Antenna power pattern for Novosibirsk detector for the optimal angle.}
\label{fig:4}
\end{figure}

\begin{figure}
\centering
\includegraphics[width=7 cm]{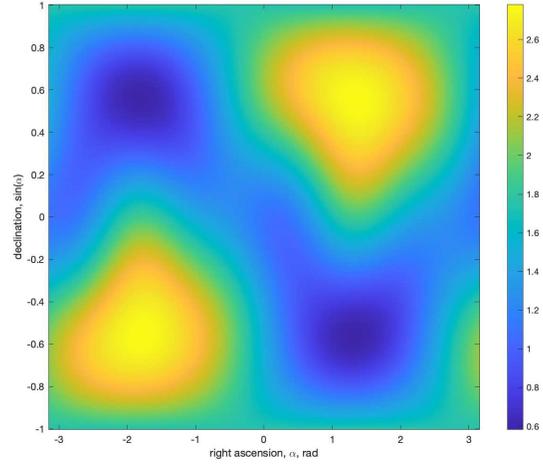}
\caption{EAN network power pattern for the optimal angle.}
\label{fig:5}
\end{figure}

\begin{figure}
\centering
\includegraphics[width=7.2 cm]{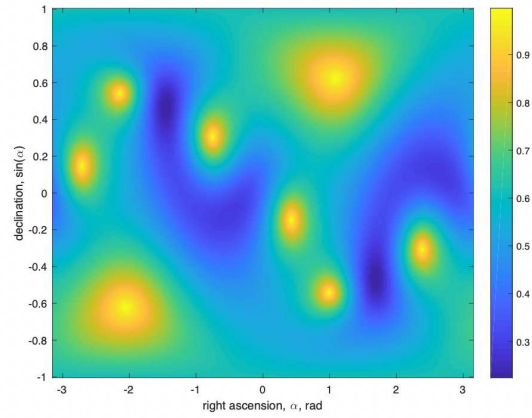}
\caption{Ratio $F_{\times}^{N}/F_{+}^{N}$ for EAN in DPF for the optimal angle.}
\label{fig:6}
\end{figure}

DN antenna power pattern is presented in Figure~\ref{fig:4}. It is a typical antenna power pattern of a single detector with two areas of effective reception on the opposite sides of celestial sphere and four areas of weak reception. The EAN network antenna power pattern  ${ P }^{ N }$ is shown in Figure~\ref{fig:5}. When compared to Figure~\ref{fig:4}, the effective reception zones are slightly enlarged, while zones of weak reception are blurred. Figure~\ref{fig:6} presents ratio $\frac {{ F }_{\times}^{ N }  }{ { F }_{+}^{ N } }$ for EAN calculated in DPF and shows eight areas, where ${ F }_{\times}^{ N }$ and ${F}_{+}^{ N }$ are almost equal so polarizations can be confidently reconstructed. Figures~\ref{fig:4}--\ref{fig:6} are constructed for the optimal angle ${ \gamma  }_{ Nsk }^{max}={ 13 }^{ \circ  } $.  

\begin{figure}
\centering
\includegraphics[width=7 cm]{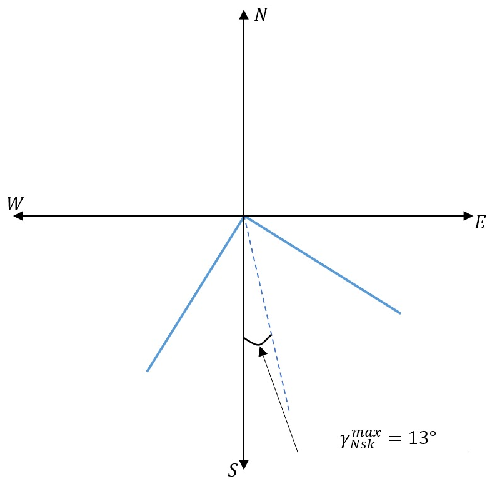}
\caption{Novosibirsk detector optimal orientation.}
\label{fig:7}
\end{figure}

 In addition, Figure~\ref{fig:7} shows the optimal orientation of the detector in Novosibirsk with respect to the cardinal directions.

Table~\ref{tab:3} shows the ratio of the criteria values calculated for the EAN network with the optimal detector angle in Novosibirsk and the LHVI network, as well as LHVK.
\begin{table}
\caption{Ratio of criteria.}
\label{tab:3}
\centering
\begin{tabular}{cccc}
\textbf{Network} & \textbf{I} & \textbf{D} & \textbf{R}\\
LHVK & 1.4 & 0.6 & 0.9 \\
LHVI & 1.2 & 0.7 & 1.0 \\
\end{tabular}
\end{table}
It can be seen that the EAN network better copes with the recovery of the GW polarization, than~the LHVK and LHVI networks, as well as being relatively effective in restoring the source parameters. The only criterion by which the EAN is inferior to the other two networks is the localization criterion. This is because all of the EAN detectors are located in Eurasia but LHVK and LHVI networks in Eurasia and North America, providing the best possible base. 

In conclusion, we would like to point out the relevance of the calculation of the EAN network efficiency for GW signals from collapsing stars. This problem has not yet found its solution with the help of the LIGO-VIRGO detector network, although the astrophysical probability of the appearance of a GW signal accompanying the collapse of stars per one galaxy is higher than for signals from their merger. If we take that a rate of events for collapsing star appearing coincides with the rate of events for supernova explosion, it turns out to be at least two orders of magnitude higher than the frequency of the binaries merger per galaxy, namely it is estimated as $0.01$ {year}$^{ -1 }$. The estimation of the expected amplitude of the GW burst from the collapse is $10^{-20}$--$10^{-22}$ for a source in the center of our galaxy \cite{b-kogan2017}, which is in the zone of current sensitivity of interferometers LIGO and VIRGO. Interest in recording the signals from the collapse is increased by the existence of parallel neutrino radiation, i.e., the possibility of implementing the multi-messenger astronomy. Apparently, the signals of both radiations contain much richer physics than the gravitational bursts (chirps) emitted by the merger of relativistic binaries~\cite{Schutz2009}  (the latter is well described already in the Newtonian theory of paragraph 3, and the fine relativistic PPN details of the chirps are not yet resolved). The interconnected temporal structure of neutrino and gravitational bursts from collapsing stars can serve as a unique indicator of the nuclear processes occurring in it \cite{b-kogan2017,Mayle1987,Janka2015}. In particular, bounds in the process of monotonic compression indicate a change in the equations of nuclear matter state with increasing density, temperature, etc. This argument is the main motive of the program to search for signals from collapsars. There are a number of scenarios that predict the structure of GW signals from the collapse~\cite{Fryer2011}. However, they all contain a hypothetical element in the model. In this respect, the calculation of criteria $ R $ and $ C $ for certain pulse structures will actually serve as a test of the corresponding models of the theory.

\vspace{6pt}

\authorcontributions{Conceptualization, V.R.; software, S.A., G.M., D.K.; formal analysis, V.R., D.K.; data~curation, G.M.; writing---original draft preparation, V.R.; writing---review and editing, S.A., G.M., D.K.; visualization, S.A.; supervision, V.R. All authors have read and agreed to the published version of the manuscript.}

\vspace{6pt}

\funding{This research was funded by  grant RFBR 19-29-11010.}

\vspace{6pt}

\acknowledgments{The authors are grateful to P. Raffai for clarifying the delicate issues related to quality assessment criteria of GW-networks. They are also grateful to the President of the Russian Gravity Society
A.A. Starobinsky and Rector of Dubna University V. D. Fursaev for the discussion of the EAN project within the framework of the meeting ''Russian Projects Mega Science in gravitation''.}

\vspace{6pt}

\conflictsofinterest{The authors declare no conflict of interest. The funders had no role in the design of the study; in the collection, analyses, or interpretation of data; in the writing of the manuscript, or in the decision to publish the results.} 

\vspace{6pt}

\abbreviations{The following abbreviations are used in this manuscript:\\

\noindent 
\begin{tabular}{@{}ll}
GW & Gravitational waves\\
BH & Black holes\\
NS & Neutron stars\\
EAN & Euro-Asian Network\\
DPF & Dominant polarization frame \\
DN & Detector in Novosibirsk\\
SNR & Signal-to-noise ratio 
\end{tabular}}


\begin{thebibliography}{999}

\bibitem{Abbott061102}
Abbott, B.P.; Abbott, R.; Abbott, T.D.; Abernathy, M.R.; Acernese, F.; Ackley, K.; Adams, C.; Adams, T.; Addesso, P.; Adhikari, R.X.; et al. Observation of Gravitational Waves from a Binary Black Hole Merger.  {\em Phys. Rev. Lett.} {\bf 2016}, {\em 116}, 061102.

\bibitem{Abbott241103}
Abbott, B.P.; Abbott, R.; Abbott, T.D.; Abernathy, M.R.; Acernese, F.; Ackley, K.; Adams, C.; Adams, T.; Addesso, P.; Adhikari, R.X.; et al. GW151226: Observation of Gravitational Waves from a 22-Solar-Mass Binary Black Hole Coalescence. {\em Phys. Rev. Lett.} {\bf 2016}, {\em 116}, 241103.

\bibitem{Abbott221101}
Abbott, B.P.; Abbott, R.; Abbott, T.D.; Acernese, F.; Ackley, K.; Adams, C.; Adams, T.; Addesso, P.; Adhikari,~R.X.; Adya, V.B.; et al. GW170104: Observation of a 50-Solar-Mass Binary Black Hole Coalescence at Redshift 0.2.  {\em Phys. Rev. Lett.} {\bf 2017}, {\em 118}, 221101.

\bibitem{Abbott141101}
Abbott, B.P.; Abbott, R.; Abbott, T.D.; Acernese, F.; Ackley, K.; Adams, C.; Adams, T.; Addesso, P.; Adhikari,~R.X.; Adya, V.B.; et al. GW170814: A Three-Detector Observation of Gravitational Waves from a Binary Black Hole Coalescence. {\em Phys. Rev. Lett.} {\bf 2017}, {\em 119}, 141101.

\bibitem{Abbott161101}
Abbott, B.P.; Abbott, R.; Abbott, T.D.; Acernese, F.; Ackley, K.; Adams, C.; Adams, T.; Addesso, P.; Adhikari,~R.X.; Adya, V.B.; et al. GW170817: Observation of Gravitational Waves from a Binary Neutron Star Inspiral.  {\em Phys. Rev. Lett.} {\bf 2017}, {\em 119}, 161101.

\bibitem{gus-ru}
Gusev, A.V.; Rudenko, V.N. Optimal Integration of the Components of the Global Network of Gravitational-Wave Antennas. {\em Mosc. Univ. Phys.} {\bf 2019}, {\em 74}, 115--123. 

\bibitem{Sathyaprakash2012}
Sathyaprakash, B.; Abernathy, M.; Acernese, F.; Ajith, P.; Allen, B.; Amaro-Seoane, P.; Andersson, N.; Aoudia,~S.; Arun, K.; Astone, P.; et al. Scientific objectives of Einstein Telescope. {\em Clas. Quant. Grav.} {\bf 2012}, {\em 29},~124013.

\bibitem{Raffai2013}
Raffai, P.; Gondan, L.;  Heng, I.S.; Kelecsenyi, N.; Logue, J.; Marka, Z.; Marka, S. Optimal networks of future gravitational-wave telescopes . {\em Clas. Quant. Grav.} {\bf 2013}, {\em 30}, 155004.

\bibitem{Raffai2015}
Hu, Y.M.; Raffai, P.; Gondan, L.; Heng, I.S.; Kelecsenyi, N.; Hendry, M.; Marka, Z.; Marka, S. Global~optimization for future gravitational wave detector sites.  {\em Clas. Quant. Grav.} {\bf 2015}, {\em 32}, 105010.

\bibitem{Novosibirsk}
Russian National Projects of ''Megasciense'' Class in Gravitational Physics.
Available online: \textit{http://gravity-conf.uni-dubna.tilda.ws/} (accessed on 20 June 2020).

\bibitem{Arnaud2002}
Arnaud, N.; Barsuglia, M.; Bizouard, M.A.; Canitrot, P.; Cavalier, F.; Davier, M.; Hello, P.; Pradier,~T. Detection~in coincidence of gravitational wave bursts with a network of interferometric detectors: Geometric~acceptance and timing.
{\em Phys. Rev. D} {\bf 2002}, {\em 64}, 042004.

\bibitem{Akutsu2018}
Akutsu, T.; Ando, M.; Araki, S.; Araya, A.; Arima, T.; Aritomi, N.; Asada, H.; Aso, Y.; Atsuta, S.; Awai, K.; et~al. Construction of KAGRA: An underground gravitational-wave observatory. {\em Prog. Theor. Exp. Phys.} {\bf 2002}, {\em 2018}, 013F01.

\bibitem{India2019}
Souradeep, T. LIGO-India: Origins and site search.  In Proceedings of the LIMMA-2019, Dukes Retreat, Khandala, India, 15--18 January 2019.

\bibitem{Schutz2009}
Sathyaprakash, B.S.; Schutz, B.F. Physics, Astrophysics and Cosmology with Gravitational Waves. {\em Living~Rev. Relativ.} {\bf 2009}, {\em 12}, 2.

\bibitem{Schutz2011}
Schutz, B.F. Networks of gravitational wave detectors and three figures of merit. {\em Clas. Quant. Grav.} {\bf 2011}, {\em 28},~125023.

\bibitem{JaranowskiSchutz1998}
Jaronowski, P.; Krolak, A.; Schutz, B.F. Data analysis of gravitational-wave signals from spinning neutron stars: The signal and its detection.  {\em Phys. Rev. D} {\bf 1998}, {\em 58}, 063001.


\bibitem{Bonazolla1996}
Bonazolla, S.; Gourgoulhon, E. Gravitational waves from pulsars: Emission by the magnetic-field-induced distortion. {\em Astron. Astrophys.} {\bf 1996}, {\em 312}, 675--690.

\bibitem{Freise2009}
Freise, A.; Chelkowski, S.; Hild, S.; Del Pozzo, W.; Perreca, A.; Vecchio, A. Triple Michelson interferometer for a third-generation gravitational wave detector. {\em Clas. Quant. Grav.} {\bf 2009}, {\em 26}, 085012.

\bibitem{Finn2001}
Finn, L.S. Aperture synthesis for gravitational-wave data analysis: Deterministic sources. {\em Phys. Rev. D} {\bf 2001}, {\em 63}, 102001.

\bibitem{MKW}
Misner, C.W.; Thorne, K.S.; Wheeler, J.A. {\em Gravitation v. 3}; W. H. Freeman and Company: San Francisco, CA, USA, 1973.

\bibitem{Finn1993}
Finn, L.S.; Chernoff, D.F. Observing binary inspiral in gravitational radiation: One interferometer. {\em Phys.~Rev.~D}{\bf 1993}, {\em 47}, 2198.

\bibitem{BZR}
Braginskii, V.B.; Zeldovich, Y.B.; Rudenko, V.N. Reception of gravitational radiation of extraterrestial origin. {\em Sov. Phys. JETP} {\bf 1969}, {\em 10}, 280--283. 

\bibitem{Cutler1994}
Cutler, C.; Flanagan, E.E. Gravitational waves from merging compact binaries: How accurately can one extract the binary’s parameters from the inspiral waveform? {\em Phys. Rev. D} {\bf 1994} {\em 49}, 2658.

\bibitem{Klimenko2011}
Klimenko, S.; Vedovato, G.; Drago, M.; Mazzolo, G.; Mitselmakher, G.; Pankow, C.; Prodi, G.; Re, V.; Salemi,~F.; Yakushin, I. Localization of gravitational wave sources with networks of advanced detectors. {\em Phys. Rev. D} {\bf 2011}, {\em 83}, 102001.

\bibitem{Drago2010}
Drago, M. Search for Transient Gravitational Wave Signals with a Known Waveform in the Ligo Virgo Network of Interferometric Detectors Using a Fully Coherent Algorithm.  Ph.D. Thesis, Universita' Degli Studi di Padova, Padova, Italy
, 2010.

\bibitem{Fairhurst2011}
Fairhurst, S. Source localization with an advanced gravitational wave detector network. {\em Clas. Quant. Grav.} {\bf 2011}, {\em 28}, 105021.

\bibitem{O1data}
Public LIGO Document Control Centre. Available online: \textit{https://dcc.ligo.org/LIGO-T0900288/public} (accessed on 15 June 2020).

\bibitem{Kocsis2007}
Kocsis, B.; Haiman, Z.; Menou, K.; Frei, Z. Premerger localization of gravitational-wave standard sirens with LISA:
Harmonic mode decomposition. {\em Phys. Rev. D} {\bf 2007}, {\em 76}, 022003.

\bibitem{LIGOReport2010}
Report of the Committee to Compare the Scientific Cases for Two Gravitational-Wave Detector Networks: (AHLV) Australia, Hanford, Livingston, VIRGO; and (HHLV) Two Detectors at Hanford, One at Livingston, and VIRGO. Available online: \textit{https://dcc.ligo.org/public/0011/T1000251/001/} (accessed on 28 June 2020).

\bibitem{b-kogan2017}
Bisnovatyi-Kogan, G.S.; Moiseenko, S.G. Gravitational waves and core-collapse supernovae. {\em Phys. Usp.} {\bf 2017}, {\em 60}, 843. 

\bibitem{Mayle1987} 
Mayle, R.; Wilson, J.R.; Schramm, D.N. Neutrinos from gravitational collapse. {\em Astrophys. J.} {\bf 1987}, {\em 318}, 288--306. 
\newpage
\bibitem{Janka2015} 
Melson, T.; Janka, H.-T.; Bollig, R.; Hanke, F.; Marek, A.; Müller, B. Neutrino-driven explosion of a 20~solar-mass star in three dimensions enabled by strange-quark contributions to neutrino-nucleon scattering. {\em Astrophys. J. Lett.} {\bf 2015}, {\em 808}, L42. 

\bibitem{Fryer2011}
Fryer, C.L.; New, K.C.B. Gravitational Waves from Gravitational Collapse. {\em Living Rev. Relativ.} {\bf 2011}, {\em 14}, 1.







\end{thebibliography}
\end{document}